\documentstyle[12pt]{article}

\def\eps{\varepsilon}
\newcommand{\pa}[2]{\fracd{\partial #1}{\partial #2}}
\newcommand{\pdd}[2]{\fracd{\partial ^{2}#1}{\partial #2^{2}}}
\newcommand{\fracd}[2]{\displaystyle\frac{#1}{#2}}

\input{epsf}

\begin{document}
\setlength{\baselineskip}{0.7cm}

\textheight22.5cm
\textwidth16cm
\setlength{\oddsidemargin}{0cm}
\setlength{\unitlength}{.5cm}
\setlength{\topmargin}{0cm}
\setlength{\headheight}{0cm}
\setlength{\headsep}{0cm}
\setlength{\footheight}{1.5cm}
\setlength{\footskip}{2.3cm}
\footnotesep 10pt

\begin{center}
{\Large {\bf GLOBAL MODES FOR THE COMPLEX GINZBURG-LANDAU EQUATION}} \\[.4cm]
St\'ephane Le Diz\`es\footnote{Present address: IRPHE, 12 Avenue G\'en\'eral
Leclerc, F-13003 Marseille, France}  \\[.2cm]
Laboratoire d'Hydrodynamique (LadHyX), Ecole polytechnique,
\\[-0.1cm]
F-91128 Palaiseau cedex, France.
\end{center}

\noindent {\Large {\bf  Abstract}}

\setlength{\baselineskip}{0.55cm}
Linear global modes, which are time-harmonic solutions
with vanishing boundary conditions, are analysed in the
context of the complex Ginzburg-Landau equation with
slowly varying coefficients in doubly infinite domains.
The most unstable modes are shown to be characterized
by the geometry of their Stokes line network: they
are found to generically correspond to a configuration
with two turning points issued from opposite sides
of the real axis which are either merged
or connected by a common Stokes line.
A region of local absolute instability
is also demonstrated to be a necessary condition for the existence
of unstable global modes.

\setlength{\baselineskip}{0.65cm}

\section{Introduction}

Wakes, jets, counterflow mixing layers are well-known to
exhibit under certain flow conditions self-sustained oscillations.
It is well-established that these vortical structures
result from a Hopf bifurcation and are due to a global
destabilization of the underlying non-parallel basic flow
(Mathis {\it et al.} 1984, Monkewitz {\it et al.} 1990,
Strykowski \& Niccum 1991).
Furthermore, they seem to be connected to the appearance  of a
sufficiently large region of local absolute instability
(Monkewitz 1990).
Most theoretical investigations have so far used
 a WKBJ approach which assumes
the basic flow to be  {\bf weakly non-parallel}
in the streamwise direction.
As explained in Huerre \& Monkewitz (1990), it
permits to define local stability
properties and to link the global spatio-temporal behavior
of the perturbation in the streamwise direction to the
local dispersion relation.

The complex Ginzburg-Landau equation with slowly
varying coefficients  appears to be the
simplest model providing a scenario of transition
consistent with experiments (Huerre \& Monkewitz 1990).
Global modes, which are
 time-harmonic solutions satisfying vanishing boundary
conditions have been analysed, either with
linear or quadratic coefficients (Chomaz {\it et al.} 1988), or
under specific
assumptions on the number of turning points involved
in the frequency selection criterion (Chomaz {\it et al.}
1991, Le Diz\`es {\it et al.} 1994).

The main objective of the present article is to obtain
a characterization of the most unstable global modes
without any assumption on the coefficients.
An upper bound for the growth rate $\Im m[\omega]$ of
the most unstable global mode is obtained  in  section  2.
In section 3, the evolution
with respect to the frequency
of separation curves -or Stokes
lines-  is analysed in the complex plane in order  to identify
most unstable global modes configurations.
Characteristic properties of global modes
and expressions satisfied by their frequency
are given in section 4.
The last section recalls the main results and
briefly discusses their
extension to the weakly nonlinear
r\'egime.

\section{Necessary conditions for the existence of
a global mode}

In the present framework, a global mode
of global frequency $\omega$ is a time-harmonic function
of the form
\begin{equation}
\Psi (x,t;\omega,\eps) = \psi (X;\omega, \eps) e^{-i\omega  t}~~,
\label{exp:Psi}
\end{equation}
where $\psi$ satisfies
the Ginzburg-Landau equation
\begin{equation}
\left[\omega  + \frac{1}{2}\omega _{kk}(X)\pdd{}{x} - i\omega
_{kk}(X)k_{0}(X) \pa{}{x} - (\frac{1}{2}\omega _{kk}(X)k_{0}^2(X)+
\omega _0(X))\right]
\psi (X;\omega ,\eps ) = 0~~,
\label{equ:psi}
\end{equation}
and vanishing boundary conditions at $+\infty$ and $-\infty$.
The coefficients $\omega_0(X)$, $\omega _{kk}(X)$ and
$k_{0}(X)$ are assumed to be
complex analytic functions
 of the slow variable
$X = \eps x$ where $\eps$ is a small positive parameter,
and  $\omega_{kk}(X)$ is
nonzero in the entire complex
plane.

The local dispersion relation associated to (\ref{equ:psi})
has the following form~:
\begin{equation}
\omega = \omega _0(X) +   \frac{\omega _{kk}(X)}{2}(k- k_{0}(X))^2~~.
\label{equ:dispersion}
\end{equation}
Causality  implies
that  there exists a local maximum growth rate for the perturbation
of  the basic flow,
which means that the quantity
$\omega _{i,max}(X) =\max_{k\in R}(\omega_i(k,X))$ is everywhere
defined as well as its maximum over all  $X$ real denoted
$\omega _{i,max}^{max}$.
This condition  guarantees also the existence of
$\omega _{0,i}^{max}$ and  $\omega _{i,max}(\infty)$
 respectively defined as being
the  maximum of
$\omega _{0,i}(X)$ over
all $X$ real and the larger of the
two limiting values taken by $\omega _{i,max}(X)$
at $X=+\infty$ and $X=-\infty$.
Note
that the wavenumber $k=k_0(X)$ has a vanishing group velocity
$\pa{\omega}{k}(k_0)=0$. According to Bers (1983), $k_0(X)$ and
$\omega _0(X) =\omega (k_0,X)$ are indeed
the local absolute wavenumber and the local absolute frequency,
respectively. The sign of $\omega _{0,i}(X)$ characterises the
absolute/convective nature of the local instability. If
$\omega _{0,i}(X) >0$, the location $X$
is said to be (locally) absolutely unstable. It is
convectively unstable or stable, otherwise.

It is well-known that
there exist two formal solutions of (\ref{equ:psi})
which admit the following WKBJ approximations
as $\eps \rightarrow 0$ (Bender \& Orszag 1978, Wasow 1985)
\begin{equation}
\psi^{\pm} \sim A^{\pm}(X;\omega)e^{\frac{i}{\eps}\int^{X}k^{\pm}
(S;\omega )dS}~~,
\label{exp:WKBJ}
\end{equation}
where the two local wavenumbers $k^+$ and $k^-$ are defined as
\begin{equation}
k^{\pm}(X;\omega)= k_0(X) \pm \sqrt{2\frac{\omega -\omega _0(X)}{\omega
_{kk}(X)}} ~~.
\label{exp:localk}
\end{equation}
The square root can be arbitrarily defined, but for convenience,
we shall use in this section  the value of the square root of
positive imaginary part with a branch cut on the positive real
axis.

The solutions $\psi ^+$ and $\psi ^-$ are not asymptotically
valid in a full complex neighborhood of turning points
defined\footnote{Note that these points are  the branch
points of the square root and satisfy also $\omega _0(X)=\omega$.}
by
\begin{equation}
k^{+}(X;\omega) =k^{-}(X;\omega)~~.
\label{def:TP}
\end{equation}
However, they are known {\bf to be uniformly valid approximations
of solutions of (\ref{equ:psi})
in any (bounded) domain where any two points can be
connected by a path along which
$\Im m[\int^X (k^+(S;\omega) -k^-(S;\omega))dS]$ is a
differentiable and nondecreasing function}
(Wasow 1985, Fedoriuk 1987).

In the present context,
WKBJ approximations are useful only if  boundary conditions
can be applied to them, and therefore
only if they are uniformly valid
from arbitrarily large negative $X$ to arbitrarily large positive
 $X$.
Their use then requires a precise analysis
in the complex $X$ plane of the separation curves
$\Im m[\int^X (k^+(S;\omega) -k^-(S;\omega))dS]= Cst$ in order to
identify the region of uniform validity, as well as
a study of the signs of $\Im m[k^+]$ and $\Im m[k^-]$
 near infinity along the real axis to determine the behavior of each
WKBJ approximation for large $X$.

\begin{figure}[htb]
\hspace{0.4cm}\epsffile{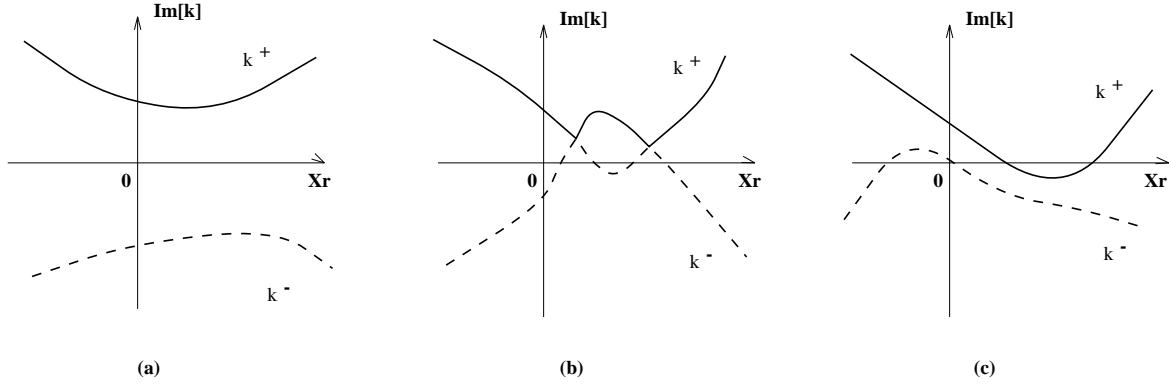}
\caption{Typical sketches of the imaginary
part of the spatial branches  $k^+(X;\omega)$ and
$k^-(X;\omega)$ versus the real location $X_r$
at a fixed complex value of $\omega$ with the square root definition
prescribed in the text. (a):
$\omega _{i,max}^{max}<\omega _i$;
(b): $\omega _{i,max}(\infty)< \omega _i \leq \omega _{i,max}^{max}$;
(c): $\max [\omega _{i,max} (\infty),\omega _{0,i}^{max}]<
\omega _i \leq \omega _{i,max}^{max}$.}
\label{fig:branchk}
\end{figure}

According to the above definitions, the signs
of the imaginary part  of both branches $k^+$ and
$k^-$ are easily determined.
For all $\omega$ such that $\omega_{i} > \omega _{i,max}^{max}$, and
all real $X$ there is no real $k$ solution of the local
dispersion relation (\ref{equ:dispersion}),
and when $\omega _i \rightarrow +\infty$,
the branches $k^+$ and $k^-$ are opposite. It follows that
for all $\omega _i > \omega _{i,max}^{max}$ and all real $X$,
$\Im m[k^+]$ and $\Im m[k^-]$
are differentiable, of constant and opposite sign
as illustrated on Figure
\ref{fig:branchk}(a).
When $\omega _i$ becomes inferior to $\omega _{i,max}^{max}$,
 $\Im m[k^+]$ and $\Im m[k^-]$  eventually change sign.
But, by definition, as long as  $\omega_{i} > \omega _{i,max} (\infty)$,
these changes do not occur near $\infty$ [Figure
\ref{fig:branchk}(b)].  Behaviors for large $X$
of both WKBJ approximations
are then invariant for all
$\omega_{i} > \omega _{i,max} (\infty)$~: $\psi ^+$ is exponentially
small for large positive real $X$ and  exponentially large
for large negative real $X$ and the opposite holds for $\psi ^-$.

When $\omega _i > \omega _{0,i}^{max}$,
the spatial branches $k^+$ and $k^-$~ satisfy
an additional property~:
 according to the definition
of the square root prescribed above, both branches are
analytic complex functions for all real $X$ and satisfy
$\Im m[k^+(X;\omega) - k^-(X;\omega)] >0$
[Figure \ref{fig:branchk}(c)].  The function
$\Im m[\int^X (k^+(S;\omega) -k^-(S;\omega))dS]$ is then a
differentiable and nondecreasing
function on the real $X$ axis, so that
$\psi ^+$ and $\psi^-$ are uniformly valid asymptotic solutions
in any interval of the real axis.

The behaviors of these two independent solutions
for large $X$  finally  imply that
no solution vanishing on both sides at $\infty$ can be constructed
\footnote{The uniform validity near infinity
of WKBJ approximations is not essential for the proof, although
it can be achieved under certain
conditions (Fedoriuk 1987).}.
In other words, all global modes have a frequency satisfying
\begin{equation}
\omega _i \leq \max [\omega _{i,max} (\infty),\omega _{0,i}^{max}] ~~.
\label{ineq:GF}
\end{equation}
In physical terms, this in particular means  that a medium which is
stable at infinity ($\omega _{i,max} (\infty) < 0$)
is globally stable if there is no region of
local absolute instability, i.e. no location where
$\omega _{0,i}(X)>0$.

Note that this result is stated in Chomaz {\it et al.} (1991)
with an incomplete proof and proved in Le~Diz\`es {\it et al.}
(1994) by another method in the case of two-turning-point
configurations.

\section{Stokes line network deformations to a global
mode configuration}

In this section, we assume that $\omega _{i,max}(\infty) <
\omega _{0,i}^{max}$ and show that the analysis of the separation
curves $\Im m[\int^X (k^+(S;\omega) -k^-(S;\omega))dS]= Cst$ in
the complex plane permits to identify the
most unstable global mode configurations
of frequency $\omega$ such that $\omega _i > \omega _{i,max}(\infty)$.
\begin{figure}[htb]
\hspace{1.5cm}\epsffile{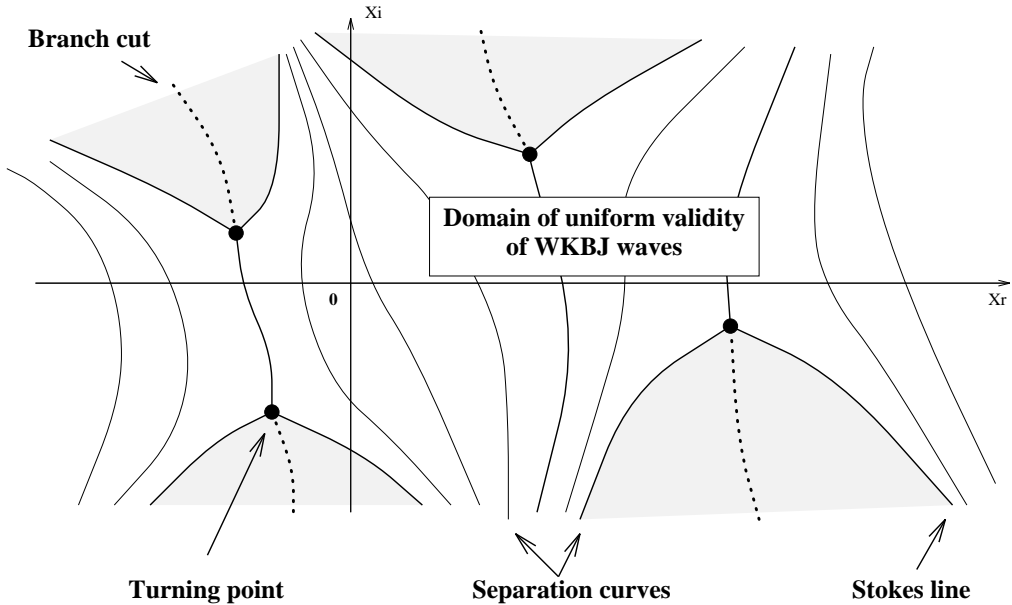}
\caption{Example of separation curve network for a frequency $\omega$
such that $\omega_i >\omega _{0,i}^{max}$.}
\label{fig:Sepa1}
\end{figure}

If  $\omega _i > \omega _{0,i}^{max}$, the geometry
close to the real axis of the separation
curves is particularly
simple~:  with
the definition of the square root prescribed above,
the real $X$ axis crosses each line
once and only once, and branch cuts and turning points are far
from the real axis.
The domain of uniform validity of WKBJ approximations can be
extended in the complex $X$ plane by following the simple procedure
that consists in deforming its frontier without cutting
any branch cut, nor twice separation curves of the same label.
The largest domain that can be obtained
is limited by the Stokes
lines\footnote{i.e. separation curves issued from a turning point}
that do not cross the real axis\footnote{A
redefinition of the square root
may be needed to maintain branch cuts in the
grey sector limited by Stokes lines
of Figure \ref{fig:Sepa1}.}
as illustrated on Figure
\ref{fig:Sepa1}.

As long as  $\omega _i > \omega _{i,max}(\infty)$,
none of the WKBJ approximations
satisfies simultaneously both
boundary conditions, as  obtained
above (see Figure \ref{fig:branchk}(b)).
Furthermore,
for these values, the geometry
of the separation curves remains unchanged near infinity
because no
turning points goes to infinity and
none of the separation
curves become
asymptotic to the real axis near $\infty$.
However, when $\omega _i$ is  smaller
than $ \omega _{0,i}^{max}$, some turning
points may have crossed the real axis and some separation
curves become twisted but the domain
of uniform validity of  WKBJ approximations may still
connect $+\infty$ and $-\infty$ on the real axis, which guarantees
the non-existence of global modes.
This fails as soon as {\bf two
turning points or two Stokes lines
which were on opposite sides of the real axis
for large $\omega _i$  collide}.
\begin{figure}[htb]
\hspace{0.1cm}\epsffile{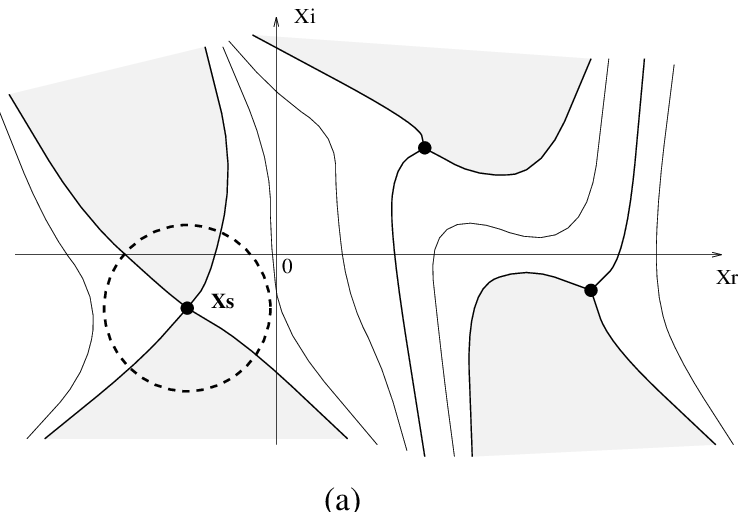}\hspace{0.6cm}\epsffile{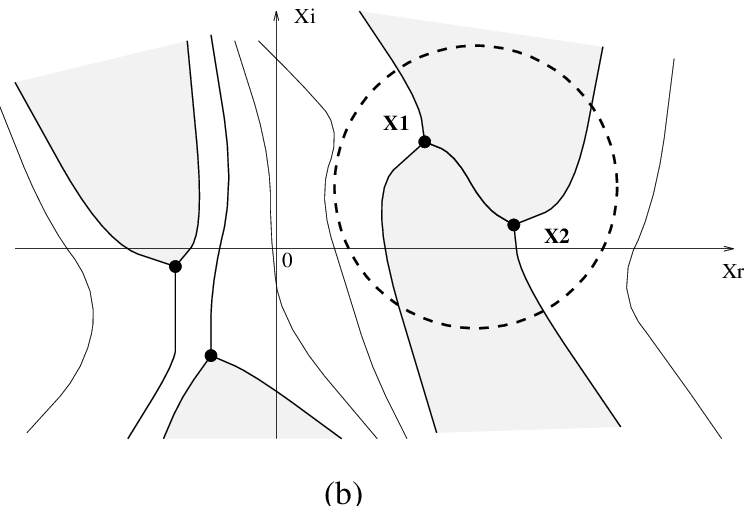}
\caption{Generic pinching configurations~;
(a): Pinching between two turning points,
(b): Pinching between two Stokes lines.}
\label{fig:Sepa3}
\end{figure}

When such a Stokes line configuration occurs,
the domain of uniform validity becomes ``pinched''
between two turning points [Figure \ref{fig:Sepa3}(a)]
or two Stokes lines [Figure \ref{fig:Sepa3}(b)]
and both infinities on the real axis
are no longer in the same domain.
The function
$\Im m[\int^X (k^+(S;\omega) -k^-(S;\omega))dS]$ becomes stationnary
at the double turning point $X_s$ in
case (a), and on the Stokes line connecting
the single turning points $X_1$ and $X_2$ in case (b).
The uniform validity of the WKBJ approximation from large negative
$X$
to large positive $X$ is then no longer guaranteed and the existence
of a global mode cannot be excluded by the former argument.

One can indeed demonstrate
that the frequency that corresponds to one of these Stokes
line networks is actually the leading-order approximation
to a discrete number
of global frequencies.
The next section gives a outline of the proof.

\section{Most unstable global modes characteristics}

The simplest proof is based
on the local analysis of turning point regions.
Since WKBJ approximations are still uniformly valid
on each side of the region where  pinching has occurred,
the existence of a global
mode is equivalent to the correct matching of  the
 WKBJ approximations prescribed by the boundary conditions
at infinity\footnote{i. e. both  WKBJ approximations subdominant
at $+\infty$ and $-\infty$, respectively.}
across either
the double turning point $X_s$ [Figure \ref{fig:Sepa3}(a)],
or the two single turning points $X_1$
and $X_2$ connected by a Stokes line [Figure \ref{fig:Sepa3}(b)].

This kind of problem is treated in several textbooks for
real configurations (see for instance
Bender \& Orszag, 1981, chap. 10). The same analysis
can easily be applied in the
complex plane and leads to the following results.
\begin{figure}[htb]
\hspace{1cm}\epsffile{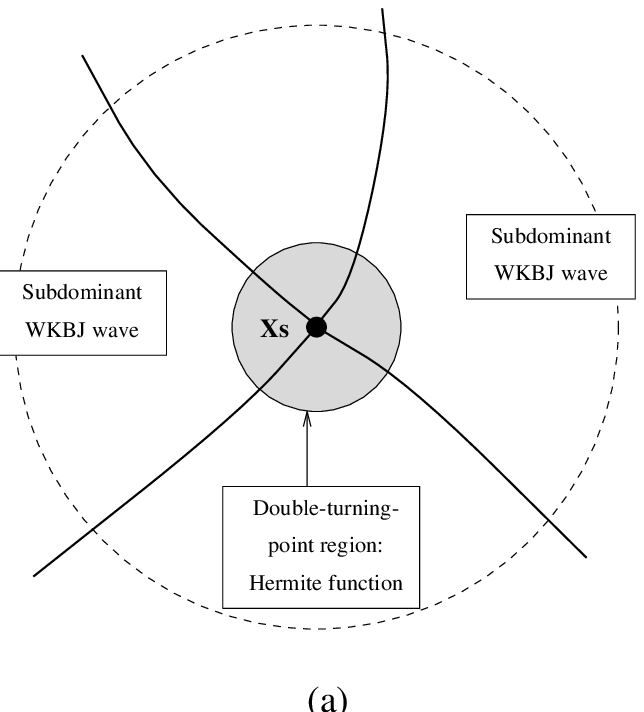}
\hspace{1.5cm}
\epsffile{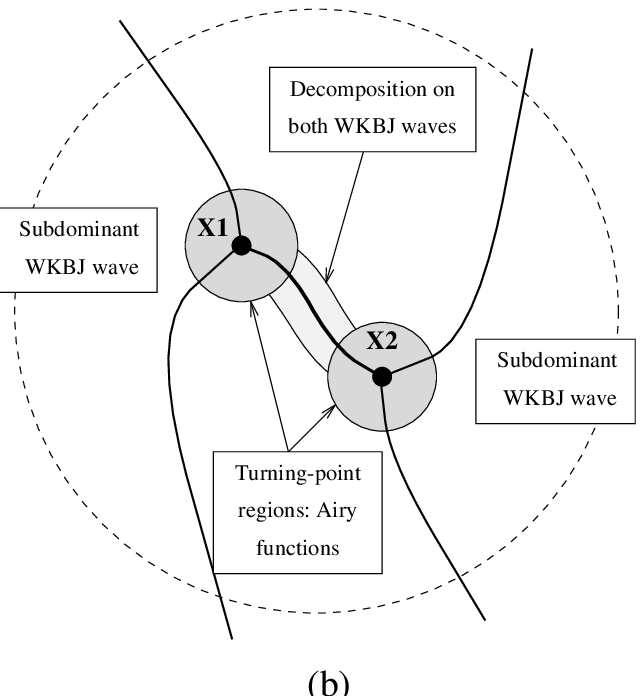}
\caption{Global modes approximations in the complex $X$ plane~;
(a)~: Double-turning-point global mode,
(b)~: Two-single-turning-points global mode.}
\label{fig:GM}
\end{figure}

For the double-turning-point configuration,
 matching is possible if and  only
if the local approximation
near the double-turning-point region
is a Hermite function [see Figure \ref{fig:GM}(a)].
This gives a discrete number of global modes
whose leading-order frequency $\omega ^{(0)}$
is the local absolute frequency taken
at the double turning point $X_s$~:
\begin{equation}
\omega ^{(0)} = \omega _{0}(X_s) ~.
\end{equation}

For the two-single-turning-point configuration,
one has to proceed to a double matching
because there are three large
separated regions [see Figure \ref{fig:GM}(b)].
Along the Stokes line connecting both turning points, any solution
admits a uniform decomposition  into both families of WKBJ waves.
But, that approximation breaks down near each turning point
where a local approximation is built in terms of Airy functions.
These local approximations can themselves be matched to
each subdominant WKBJ approximation if and only if
the following integral relation
is satisfied by the
leading
order global frequencies $\omega_n ^{(0)}$
\begin{equation}
\frac{1}{\pi\eps}\int_{X_1}^{X_2}\sqrt{2\frac{\omega _n ^{(0)}
-\omega _0(X)}{\omega _{kk}(X)}} dX \sim  n+\frac{1}{2} ~~.
\label{eigenvalue2}
\end{equation}
This relation, where $n$ is an integer
of order $1/\eps$, constitutes the leading-order
expression satisfied by a discrete number of global modes.
When $n$ is of order unity,
both turning points are
distant of $\eps$ and one recovers the double-turning-point case.
The same results can also be obtained using  a
method based on uniform approximations
as shown by Le Diz\`es {\it et al.} (1994).

\section{Discussion}

We have shown that a global mode
for the Ginzburg-Landau equation
has always
a growth rate $\omega _i$ smaller
than $\max [\omega _{i,max} (\infty),\omega _{0,i}^{max}]$.
Furthermore, the most unstable global modes of
growth rate larger than $\omega _{i,max} (\infty)$ have been
demonstrated to be described by specific
Stokes line  configurations.
They are {\bf generically}\footnote{One can point
out that nothing can in general be said
concerning more stable global modes. Furthermore, non-generic
cases corresponding
to a first pinching of more than two Stokes lines or turning points
may also give rise to a global mode which is not described by
the present analysis.}  associated
to the first pinching, when $\omega _i$ is decreased,
of two Stokes lines or turning points which were
on opposite sides of the real axis
for large $\omega _i$.

The global frequencies are discretized at order $\eps$
and satisfy the interesting property
of being only (explicitly) dependent on the local dispersion
relation (\ref{equ:dispersion}) in the region where the
pinching has occurred.
Approximations for these modes
can  easily be
obtained in the ``pinched region''
and in the domains of uniform validity
of the WKBJ approximations (white domains of Figures \ref{fig:Sepa3}),
but one must emphasize that
approximations are not necessarily known or easily accessible
everywhere on the real axis.

The extension in the weakly nonlinear r\'egime has been
studied in Chomaz {\it et al.} (1990), Le Diz\`es
{\it et al.} (1993) and Le Diz\`es (1994)
by adding a cubic nonlinearity
to the right-hand side of equation (\ref{equ:psi}).
In the simplest case where the single
 destabilized linear global mode and
its adjoint mode admit  uniform
WKBJ approximations on the real axis, the weakly nonlinear
analysis ``\`a la Stuart-Landau'' has given the surprising conclusion
that a physically acceptable solution is only obtained  under
very restrictive conditions (Le Diz\`es 1994).
It is then very likely that there only exists  a small class
of linear global modes which can effectively
saturate through a weakly nonlinear process
when destabilized. This class remains to be determined in order
to fully justify the application  to a physical problem
of the linear frequency selection criterion
obtained in the present article.

I thank Patrick Huerre and Jean-Marc Chomaz
for their helpful advice and kind encouragement.
\\[0.4cm]

\setlength{\baselineskip}{0.55cm}
\noindent {\Large {\bf Reference}}
\begin{list}{}{\setlength{\leftmargin}{0.5cm}\setlength{\listparindent}{-0.5cm}}

\item
\hspace{-0.6cm}
  Bender, C. M. \&  Orszag, S. A. ~1978~  Advanced
Mathematical Methods for Scientists and Engineers. New-York~: McGraw-Hill.

  Bers A. ~1983~  Space-time evolution of plasma
instabilities $-$absolute and convective. In {\it Handbook in Plasma
Physics} (ed. M. N. Rosenbluth and R. Z. Sagdeev), vol. 1,  451-517,
Amsterdam~: North-Holland.

  Chomaz, J. M.,  Huerre, P. \&  Redekopp, L. G. ~1988~
Bifurcations to local and global modes in spatially developing flows.
{\it Phys. Rev. Lett.} {\bf 60}, 25-28.

  Chomaz, J. M.,  Huerre, P. \&  Redekopp, L. G. ~1990~
Effect of nonlinearity and forcing on global modes. In {\it
Proceedings of the Conference on New Trends in Nonlinear
Dynamical Systems and Pattern-forming Phenomena~: The Geometry of
Nonequilibrium}
(ed. P. Coullet \&  P.
Huerre), NATO ASI Series B~: Physics, vol. 37,  259-274, New York~: Plenum
Press.

  Chomaz, J. M.,  Huerre, P. \&  Redekopp, L. G. ~1991~
A frequency selection criterion in spatially-developing flows. {\it
Stud. Appl. Math.} {\bf 84}, 119-144.

 Fedoriuk, M. ~1987~ M\'ethodes asymptotiques pour les \'equations
diff\'erentielles ordinaires lin\'eaires. Editions Mir, Moscou.

 Huerre, P. \&  Monkewitz, P. A. ~1990~ Local and global
instabilities in spatially developing flows. {\it Annu. Rev. Fluid
Mech.} {\bf 22}, 473-537.

 Le Diz\`es, S. ~1994~  Modes globaux dans les \'ecoulements
faiblement inhomog\`enes. PhD thesis (Ecole Polytechnique, France).

 Le Diz\`es, S., Huerre, P., Chomaz, J. M.
\&  Monkewitz, P. A. ~1993~
Nonlinear stability analysis of slowly-diverging
flows~: Limitations of the weakly nonlinear approach. In {\it
Proceedings of the IUTAM Symposium on Bluff-Body Wakes,
Dynamics and Instabilities} (ed. H. Eckelmann, J. M. R. Graham, P.
Huerre \&  P. A. Monkewitz), Berlin~: Springer-Verlag, 147-152.

 Le Diz\`es, S., Huerre, P., Chomaz, J. M. \&  Monkewitz,
P. A. ~1994~
Linear global modes in spatially-developing  media. Submitted
to {\it Phil. Trans. R. Soc. Lond.} A.

 Mathis, C., Provansal, M. \&  Boyer, L. ~1984~ The
B\'enard-von K\'arm\'an instability~: an experimental study near
the threshold. {\it  J. Phys. (Paris) Lett.} {\bf 45}, 483-491.

 Monkewitz, P. A. ~1990~  The role of absolute and
convective instability in predicting the behavior of fluid systems.
{\it Eur. J. Mech.} B/Fluids {\bf 9}, 395-413.

 Monkewitz, P. A., Bechert, D. W.,
Lehmann, B. \& Barsikow, B. ~1990~ Self-excited oscillations and mixing in
heated round jets. {\it J. Fluid Mech.} {\bf 213}, 611-639.

 Strykowski, P. J. \& Niccum, D. L. ~1991~ The stability of
countercurrent mixing layers in
circular jets. {\it J.	 Fluid Mech.} {\bf 227}, 309-343.

 Wasow, W. ~1985~   Linear turning point theory.
New-York~: Springer-Verlag.

\end{list}

\end{document}